\documentclass[paper,notoc]{JHEP3}

\usepackage{epsfig}

\newcommand{\beq}{\begin{equation}}
\newcommand{\eeq}{\end{equation}}

\newcommand{\bea}{\begin{eqnarray}}
\newcommand{\eea}{\end{eqnarray}}
\newcommand{\nn}{\nonumber}

\def\eqn#1{Eq.~(\ref{#1})}
\def\eqns#1#2{Eqs.~(\ref{#1}) and~(\ref{#2})}
\def\eqnss#1#2{Eqs.~(\ref{#1})-(\ref{#2})}

\def\sec#1{Section~{\ref{#1}}}

\newcommand\fverb{\setbox\pippobox=\hbox\bgroup\verb}
\newcommand\fverbdo{\egroup\medskip\noindent%
                        \fbox{\unhbox\pippobox}\ }
\newcommand\fverbit{\egroup\item[\fbox{\unhbox\pippobox}]}
\newbox\pippobox

\def\ord{{\cal O} }
\def\cM{{\cal M}}
\def\eps{\epsilon}

\def\cg{c_\Gamma}
\newcommand\sss{\scriptscriptstyle}
\newcommand\as{\alpha_{\sss S}} 
\newcommand\gs{g_{\sss S}}
\def\tgs{\tilde\gs}

\title{Testing high-energy factorization beyond the next-to-leading-logarithmic 
accuracy}

\author{Vittorio Del Duca\footnote{On leave from INFN, Sezione di Torino, Italy}\\ 
Istituto Nazionale di Fisica Nucleare\\
Laboratori Nazionali di Frascati\\
00044 Frascati (Roma), Italy\\
        E-mail: \email{delduca@lnf.infn.it}}

\author{E.~W.~N.~Glover\\
Institute for Particle Physics Phenomenology, 
University of Durham\\ Durham, DH1 3LE, U.K.\\
E-mail: \email{E.W.N.Glover@durham.ac.uk}}

\abstract{By taking the high-energy limit of the two-loop and the
three-loop four-point amplitudes in
the maximally supersymmetric $N=4$ Yang-Mill theory (MSYM),
we test the validity of the loop expansion of the high-energy amplitude, 
beyond the next-to-leading-logarithmic (NLL) accuracy.
We compute the three-loop Regge trajectory, and the 
two-loop and three-loop coefficient functions. These quantities are relevant
for the BFKL evolution beyond NLL, as well as building MSYM two-loop 
and three-loop amplitudes with many legs in the high-energy 
limit, which in turn may be used as a powerful check of the evaluation
of the corresponding exact amplitudes.}

\preprint{~IPPP/08/13}

\begin{document}

\section{Introduction}
\label{sec:intro}

In the high-energy limit (HEL) $s\gg |t|$, any scattering process is dominated
by the exchange of the highest-spin particle in the crossed channel. 
Thus, in perturbative QCD the leading contribution in powers of $s/t$
to any scattering process comes from gluon exchange in the $t$ channel.
Building upon this fact, the BFKL theory models
strong-interaction processes with two large and disparate scales,
by resumming the radiative corrections to parton-parton
scattering. This is achieved to leading logarithmic (LL) accuracy, in
$\ln(s/|t|)$, through the BFKL 
equation~\cite{Kuraev:1976ge,Kuraev:1977fs,Balitsky:1978ic}, 
i.e. a two-dimensional
integral equation which describes the evolution of the $t$-channel
gluon propagator in transverse momentum space and moment space.
The integral equation is obtained by computing the one-loop LL
corrections to the gluon exchange in the $t$ channel. They are formed
by a real correction, the emission of a gluon along the 
ladder~\cite{Grisaru:1973vw,Grisaru:1974cf,Lipatov:1976zz}, and the leading virtual contribution
of the gluon loop, the one-loop Regge trajectory~\cite{Kuraev:1976ge}.
The next-to-leading-logarithmic (NLL) corrections to the BFKL equation
have been computed as well~\cite{Fadin:1998py,Ciafaloni:1998gs}. 
The virtual part of the NLL kernel is provided by the two-loop
trajectory~\cite{Fadin:1995xg,Fadin:1995km,Fadin:1996tb,Blumlein:1998ib,DelDuca:2001gu}, 
whose evaluation is based upon the assumption of Regge factorization of the
scattering amplitudes beyond leading logarithmic accuracy.

Recently, Bern, Dixon and Smirnov (BDS) have proposed an ansatz~\cite{Bern:2005iz}  
for the $l$-loop 
$n$-gluon scattering amplitude in the maximally supersymmetric $N=4$ 
Yang-Mill theory (MSYM), with the maximally-helicity violating (MHV) configuration and for 
arbitrary $l$ and $n$. In Ref.~\cite{Bern:2005iz}, an analytic expression
for the exact three-loop four-gluon amplitude, as well as iterative relations
for the $n$-point MHV MSYM amplitudes, have also been provided. 
Using the results of Ref.~\cite{Bern:2005iz}, we test
the high-energy factorization of the gluon-gluon scattering process.
The paper is organised as follows:
in \sec{sec:factor}, we analyse high-energy factorization at three-loop
accuracy in perturbative QCD; in \sec{sec:msym}, we take the exact two-loop
and three-loop four-point MSYM amplitudes in the HEL, and 
derive the three-loop Regge trajectory, 
as well as the two-loop and three-loop coefficient functions;
in \sec{sec:bds}, we make contact between these quantities,
the BDS ansatz and the iterative relations for the MHV MSYM amplitudes;
in \sec{sec:concl}, we sketch an outlook for the BFKL evolution in MSYM
beyond NLL accuracy.

\section{High-energy factorization}
\label{sec:factor}

In perturbative QCD, the simplest process is parton-parton scattering, 
which in the HEL occurs through gluon exchange in the $t$ channel.
It is convenient to focus on
gluon-gluon scattering. In the HEL, the tree-level amplitude for
$g_a\, g_b\to g_{a'}\,g_{b'}$ may be written as \cite{Kuraev:1976ge}, 
\begin{equation}
\cM^{(0)}_4 = 2  s
\left[i\, \gs\, f^{aca'}\, C^{(0)}(p_a,p_{a'}) \right]
{1\over t} \left[i\, \gs\, f^{bcb'}\, C^{(0)}(p_b,p_{b'}) \right]\, ,\label{elas}
\end{equation}
where $a, a', b, b'$ represent the colours of the scattering gluons.
The gluon coefficient functions $C^{(0)}$, 
which yield the LO gluon impact factors, 
are given in Ref.~\cite{Kuraev:1976ge} in terms of their spin structure 
and in Ref.~\cite{DelDuca:1995zy,DelDuca:1996km} at fixed
helicities of the external gluons. 

The colour decomposition of the tree-level $n$-gluon amplitude
in a helicity basis is~\cite{Mangano:1990by}
\begin{equation}
\cM_n^{(0)} = 2^{n/2}\, g^{n-2}\, \sum_{S_n/Z_n} {\rm tr}(T^{d_{\sigma(1)}} 
\cdots
T^{d_{\sigma(n)}}) \, m_n^{(0)}(p_{\sigma(1)},\nu_{\sigma(1)};...;
p_{\sigma(n)},\nu_{\sigma(n)})\, ,\label{one}
\end{equation}
where $d_1,..., d_n$, and $\nu_1,..., \nu_n$ are
respectively the colours and the
polarizations of the gluons, the $T$'s are the colour 
matrices\footnote{We use the normalization
${\rm tr}(T^c T^d) = \delta^{cd}/2$,
although it is immaterial in what follows.} in the
fundamental representation of SU($N$) and the sum is over the noncyclic
permutations $S_n/Z_n$ of the set $[1,...,n]$. We take
all the momenta as outgoing, and consider the MHV
configurations $(-,-,+,...,+)$ for which the tree-level gauge-invariant colour-stripped
sub-amplitudes, $m_n^{(0)}(p_1,\nu_1; ...; p_n,\nu_n)$, assume the form
\begin{equation}
m_n^{(0)}(-,-,+,...,+) = {\langle p_i p_j\rangle^4\over
\langle p_1 p_2\rangle \cdots\langle p_{n-1} p_n\rangle 
\langle p_n p_1\rangle}\, ,\label{two}
\end{equation}
where $i$ and $j$ are the gluons of negative helicity. In gluon-gluon
scattering, only the helicity configurations $(-,-,+,+)$ occur at tree level,
and out of the six possible colour configurations only four are leading
in the HEL~\cite{DelDuca:1993pp}. They are 
those corresponding to $s$-channel helicity conservation. For instance,
let us label the gluons clockwise and consider the helicity configuration 
$(b-,a-,a'+,b'+)$, with $a$ and $b$ incoming and $a'$ and $b'$ outgoing,
then the four leading colour configurations are 
$(b,a,a',b')$, $(b,b',a',a)$, $(b,b',a,a')$, $(b,a',a,b')$. The latter two,
corresponding to the helicity ordering $(-,+,-,+)$, can be obtained from the
former two, corresponding to the helicity ordering $(-,-,+,+)$, by 
$s\leftrightarrow u$ channel exchange. In addition, in the HEL
$m_4^{(0)}(-,+,-,+) = - m_4^{(0)}(-,-,+,+)$, thus the different colour configurations
contribute to \eqn{one} with alternating signs, in such a way that the
traces of $T$ matrices combine to form the structure constants of
\eqn{elas}~\cite{DelDuca:1995zy}.

The virtual radiative corrections to eq.~(\ref{elas}) in
LL approximation are obtained, to all orders
in $\as$, by replacing \cite{Kuraev:1976ge}
\begin{equation}
{1\over t} \to {1\over t} 
\left({s\over -t}\right)^{\alpha(t)}\, ,\label{sud}
\end{equation}
in eq.~(\ref{elas}), where $\alpha(t)$ can be written in
dimensional regularization in $d=4-2\epsilon$ dimensions as
\begin{equation}
\alpha(t) = \gs^2\, c_{\Gamma}\,  
\left(\mu^2\over -t\right)^{\epsilon} \, N\, {2\over\epsilon}
,\label{alph}
\end{equation}
with $N$ colours, and
\begin{equation}
c_{\Gamma} = {1\over (4\pi)^{2-\epsilon}}\, {\Gamma(1+\epsilon)\,
\Gamma^2(1-\epsilon)\over \Gamma(1-2\epsilon)}\, .\label{cgam}
\end{equation}
The fact that higher order corrections to gluon exchange in the $t$ channel can
be accounted for by dressing the gluon propagator with the exponential of
\eqn{sud} is called the gluon reggeization, with $\alpha(t)$
the Regge trajectory. In order to go beyond 
the LL approximation, we need a prescription that 
disentangles the virtual corrections to the coefficient functions
in \eqn{elas} from those that reggeize the gluon (\ref{sud}). 
The prescription for doing so is supplied by
the general form of the high-energy amplitude for gluon-gluon 
scattering which arises from a single reggeized gluon exchanged in the crossed 
channel~\cite{Fadin:1993wh}
\beq
\cM_4 = s
\left[i\, \gs\, f^{aca'}\, C(p_a,p_{a'}) \right]
{1\over t} \left[\left({-s\over -t}\right)^{\alpha(t)} +
\left({s\over -t}\right)^{\alpha(t)}  \right]
\left[i\, \gs\, f^{bcb'}\, C(p_b,p_{b'}) \right]. \label{elasbg}
\eeq
The first Regge trajectory in \eqn{elasbg} corresponds to the $s$-channel
physical region; the second to the $u$ channel. The $s$ and $u$
channels have different analytic properties, however \eqn{elasbg}
presumes that the corresponding sub-amplitudes still differ only by
a sign $m_4(-,+,-,+) = - m_4(-,-,+,+)$.  It is   
clear that the equals
sign in \eqn{elasbg} cannot be expected to hold strictly, but, in fact,
holds only up to NLL accuracy~\cite{DelDuca:1998kx}. However, we show that,
at the colour-stripped amplitude level, 
the following high-energy prescription is valid beyond NLL accuracy, 
\beq
m_4(-,-,+,+) \equiv m_4^s = s
\left[\gs\, C(p_a,p_{a'}) \right]
{1\over t} \left({-s\over -t}\right)^{\alpha(t)}
\left[\gs\, C(p_b,p_{b'}) \right], \label{elasschan}
\eeq
in the $s$-channel physical region, and similarly
\beq
m_4(-,+,-,+) \equiv m_4^u = s
\left[\gs\, C(p_a,p_{a'}) \right]
{1\over t} \left({s\over -t}\right)^{\alpha(t)}
\left[\gs\, C(p_b,p_{b'}) \right], \label{elasuchan}
\eeq
in the $u$-channel physical region. In \eqnss{elasbg}{elasuchan},
the gluon Regge trajectory has the perturbative expansion,
\begin{equation}
\alpha(t) = \tilde\gs^2(t) \alpha^{(1)} + 
\tgs^4(t) \alpha^{(2)} + \tgs^6(t) \alpha^{(3)} + \ord(\tgs^8)\,
,\label{alphb}
\end{equation}
with the rescaled coupling
\beq
\tgs^2(t) = \gs^2 \cg \left({\mu^2\over -t}\right)^{\eps}\, ,\label{rescal}
\eeq
and with $\alpha^{(1)}=2N/\epsilon$ given in \eqn{alph}.
The coefficient functions $C$ can be written as
\begin{equation}
C = C^{(0)}(1 + \tgs^2(t) C^{(1)} + \tgs^4(t) C^{(2)} + 
\tgs^6(t) C^{(3)}) + \ord(\tgs^8)\, .
\label{fullv}
\end{equation}
Because the coefficient function $C$ is real (up to overall complex phases in
$C^{(0)}$ induced by the complex-valued helicity bases),
\eqnss{elasbg}{elasuchan} imply that the imaginary part of the amplitude
comes entirely from the Regge trajectory in the $s$ channel, according to the
usual prescription $\ln(-s) = \ln(s) - i\pi$, for $s > 0$.

The expansion of \eqns{elasschan}{elasuchan} can be written as,
\beq
m_4^i = m_4^{i(0)} \left( 1 +
\tgs^2\ m_4^{i(1)} + \tgs^4 m_4^{i(2)} + \tgs^6 m_4^{i(3)}
+ \ord(\tgs^8) \right)\, ,\label{elasexpand}
\eeq
with $i=s, u$. The one-loop coefficients of \eqn{elasexpand} are, 
\bea
m_4^{u(1)} &=& \alpha^{(1)} L +\ 2 C^{(1)}\, ,\nn\\
m_4^{s(1)} &=& m_4^{u(1)} - i\pi \alpha^{(1)}\, .\label{exp1loop}
\eea
with $L = \ln\left({s\over -t}\right)$.
The one-loop trajectory, $\alpha^{(1)}$, is universal, 
{\it i.e.} it is independent 
of the type of parton undergoing the high-energy scattering process. 
It is also independent of the infrared (IR) 
regularisation scheme. Conversely, the one-loop coefficient 
function, $C^{(1)}$, is process and IR-scheme dependent.
$C^{(1)}$ was computed in conventional dimensional 
regularization (CDR)/'t-Hooft-Veltman (HV) schemes in 
Ref.~\cite{Fadin:1993wh,DelDuca:1998kx,Fadin:1992zt,Fadin:1993qb,Bern:1998sc}, 
and in the dimensional reduction scheme (DRED) 
in Ref.~\cite{DelDuca:1998kx,Bern:1998sc}.

The two-loop coefficients of \eqn{elasexpand} are,
\bea
m_4^{u(2)} &=& 
{1\over 2} \left(\alpha^{(1)}\right)^2 L^2 \nn\\
&+& \left( \alpha^{(2)} + 2\, C^{(1)} \alpha^{(1)} \right)\ L 
\label{exp2loopu}\\
&+& 2\, C^{(2)} + \left(C^{(1)}\right)^2\, ,\nn\\ 
m_4^{s(2)} &=& m_4^{u(2)} - {\pi^2\over 2} \left(\alpha^{(1)}\right)^2 \nn\\
&-& i\pi \left[ \left(\alpha^{(1)}\right)^2 L
+ \alpha^{(2)} + 2\, C^{(1)} \alpha^{(1)} \right]
.\label{exp2loops}
\eea
The two-loop trajectory, $\alpha^{(2)}$, was computed in the CDR scheme in
Ref.~\cite{Fadin:1995xg,Fadin:1995km,Fadin:1996tb,Blumlein:1998ib,DelDuca:2001gu}. We note that the coefficients of the double and the real parts of the 
single logarithms are
the same in the $s$ and the $u$ channel. Therefore, it is correct to use 
\eqn{elasbg} at NLL accuracy. However, the real part of the constant term 
is channel dependent, and thus care must be used in extracting the two-loop
coefficient function from it\footnote{In Ref.~\cite{DelDuca:2001gu}, the 
constant terms for the gluon-gluon, quark-quark and gluon-quark scattering
processes were evaluated by projecting the two-loop amplitude on the
tree amplitude, which entails averaging over the $s$ and $u$ channels.
However, an apparent discrepancy in the factorization between the quark
and the gluon amplitudes prevented any conclusions from being drawn 
about the two-loop coefficient functions.}.

The three-loop coefficients of \eqn{elasexpand} are,
\bea
m_4^{u(3)} &=& 
{1\over 3!} \left(\alpha^{(1)}\right)^3 L^3 \nn\\
&+& \alpha^{(1)} \left( \alpha^{(2)} + C^{(1)} \alpha^{(1)} \right)
L^2 \label{exp3loopu}\\
&+& \left[ \alpha^{(3)} + 2\, \alpha^{(2)} C^{(1)} 
+ \alpha^{(1)} \left( 2\, C^{(2)} + \left(C^{(1)}\right)^2 \right) \right]
L \nn\\
&+& 2\, C^{(3)} + 2\, C^{(2)} C^{(1)}\, ,\nn\\
m_4^{s(3)} &=& m_4^{u(3)} - {\pi^2\over 2} \left(\alpha^{(1)}\right)^3\ L 
- \pi^2
\alpha^{(1)} \left( \alpha^{(2)} + C^{(1)} \alpha^{(1)} \right) \nn\\ &-& 
i\pi \left[ {\left(\alpha^{(1)}\right)^3\over 2} L^2
+ 2\, \alpha^{(1)} \left( \alpha^{(2)} + C^{(1)} \alpha^{(1)} \right)\ L 
\right. \label{exp3loops}\\
&& \left. \quad\ + \alpha^{(3)} + 2\, \alpha^{(2)} C^{(1)} 
+ \alpha^{(1)} \left( 2\, C^{(2)} + \left(C^{(1)}\right)^2 \right) -
{\pi^2\over 3!} \left(\alpha^{(1)}\right)^3 \right]\, .\nn
\eea
We note that the coefficients of the triple and the real part of the 
double logarithms are
the same in both $s$ and $u$ channels, thus maintaining the correctness of 
\eqn{elasbg} at NLL accuracy. However, although the three-loop trajectory is
universal, the coefficient of the single logarithm, and thus the extraction 
of the three-loop trajectory, depends on the channel under consideration.

\section{The MSYM amplitudes}
\label{sec:msym}

The QCD four-point parton-parton scattering amplitudes are known at two-loop 
accuracy~~\cite{Anastasiou:2001sv,Anastasiou:2001kg,Anastasiou:2001ue,Glover:2001af,Bern:2002tk,Anastasiou:2002zn,Bern:2003ck,Glover:2003cm,Glover:2004si}.
In QCD, the gluon loop can be decomposed into a MSYM multiplet,
an $N= 1$ chiral multiplet, and a complex scalar. The MSYM contribution is the
simplest to evaluate, and captures most of the IR behaviour of the full gluon
loop. In fact within MSYM, the gluon-gluon scattering amplitude is known  at three
loops analytically~\cite{Bern:2005iz} and at four loops through $\ord(1/\eps)$
numerically~\cite{Bern:2006ew}.  Bern, Dixon and Smirnov have proposed an 
ansatz~\cite{Bern:2005iz} 
for the MHV $l$-loop $n$-gluon  scattering amplitude for arbitrary $l$ and $n$. 
The ansatz agrees with
the direct evaluation of the three-loop four-point amplitude\footnote{
The BDS ansatz was first predicted to fail by Alday and Maldacena~\cite{Alday:2007he},
for amplitudes with a large number of gluons in the strong-coupling limit.
They claimed that the finite pieces of the two-loop six gluon amplitude would be incorrectly 
determined. This prediction was backed up by Drummond et al.~\cite{Drummond:2007bm}, who
considered the finite contribution in the dual theory~\cite{Alday:2007hr} by computing the
hexagonal light-like Wilson loop at two loops.  The conclusion was that either the BDS ansatz
is wrong, or the equivalence between Wilson loops and scattering amplitudes does not work at 
two loops. Recent numerical results for the finite part of the MHV six-gluon amplitude in 
MSYM~\cite{Bern:2008ap} have confirmed the equivalence with the finite part of the light-like 
hexagon Wilson loop~\cite{Drummond:2008aq} thereby disproving the BDS ansatz.
Furthermore, the analytic structure of the two-loop six-gluon amplitude in the multi-Regge 
kinematics is expected to be at odds with the structure of the BDS 
ansatz~\cite{Bartels:2008ce}.}
In addition, in Ref.~\cite{Drummond:2007aua,Naculich:2007ub} it has been shown
that in the HEL the BDS ansatz for the four-point amplitude exhibits the Regge
behaviour of \eqns{elasschan}{elasuchan}. Thus, we shall use the direct
evaluation of the three-loop amplitude and the BDS ansatz to derive the
relevant quantities for the three-loop high-energy factorization in MSYM.

We start with the HEL of the QCD one-loop colour-stripped
amplitude for gluon-gluon scattering, which is known to all orders in 
$\eps$~\cite{Bern:1998sc}, and use the maximal trascendentality 
principle~\cite{Kotikov:2002ab} to select the MSYM contribution.
Using the conventions of \eqn{elasexpand} and rescaling the 
coupling (\ref{rescal}) as 
\beq
{\bar\gs}^2(t) = \tgs^2(t) N\, ,\label{eq:nrescale} 
\eeq
we obtain
\bea
m_{4_{MSYM}}^{u(1)}
&=& 2\, \left(
{\psi(1+\eps) - 2\psi(-\eps) + \psi(1)\over\eps} + {L\over \eps}\right)
\label{eq:amp4alleps}\\
&=& -{4\over \eps^2} + {2 L\over \eps} +
\pi^2 + 2 \zeta_3 \eps + {\pi^4\over 15}\eps^2 + 2 \zeta_5 \eps^3
+ {2\pi^6\over 315}\eps^4 + \ord(\eps^5)\, ,\nn\\
m_{4_{MSYM}}^{s(1)} &=& m_{4_{MSYM}}^{u(1)} - {2i\pi\over\eps}\, .
\eea
From \eqn{eq:amp4alleps}, we see that the MSYM one-loop trajectory is 
precisely the same
as that in QCD, \eqn{alph}, while
the MSYM one-loop coefficient function is, to all orders in $\eps$,
given by,
\beq
C^{(1)}_{MSYM} = {\psi(1+\eps) - 2\psi(-\eps) + \psi(1)\over\eps}\,
.\label{eq:ifonel}
\eeq
%

At two loops, we take the direct calculation of the exact MSYM amplitude,
given in Ref.~\cite{Bern:2005iz} to $\ord(\eps^2)$,
and evaluate the HEL in the $u$ and $s$ physical channels,
\bea
m_{4_{MSYM}}^{u(2)} &=& 
\frac{8}{\eps^4} - \frac{8 L}{\eps^3} 
+ \left( 2L^2 - {11\pi^2\over 3} \right) \frac{1}{\eps^2} 
+ \left( {5\pi^2\over 3} L - 6\zeta_3 \right)\frac{1}{\eps}
\nn\\ &+& \left( 2\zeta_3 L - {17\over 90}\pi^4 \right) 
+ \left( {2\pi^4\over 45} L + {4\pi^2\over 3}\zeta_3 - 86\zeta_5\right) \eps 
\label{eq:a42ubds}\\ 
&+& \left[ \left(6\pi^2\zeta_3 + 86\zeta_5\right) L - 94\zeta_3^2 
- {77\pi^6\over 180} \right] \eps^2 + \ord(\eps^3)\, ,\nn\\
m_{4_{MSYM}}^{s(2)} &=& m_{4_{MSYM}}^{u(2)}
+ \frac{8i\pi}{\eps^3} - \frac{2\pi^2 + 4i\pi L}{\eps^2} 
- {5i\pi^3\over 3\eps} - 2\zeta_3 i\pi \nn\\
&& \qquad\qquad - {2i\pi^5\over 45}\eps
- \left( 6\zeta_3\pi^2 + 86\zeta_5\right) i\pi \eps^2 + \ord(\eps^3)\,
.\label{eq:a42sbds}
\eea
As expected from \eqns{exp2loopu}{exp2loops}, the real parts of
$m_{4_{MSYM}}^{u(2)}$ and $m_{4_{MSYM}}^{s(2)}$ differ by the constant term
$2\pi^2/\eps^2$. Note that the average over the $s$ and $u$ channels is in
agreement, to $\ord(\eps^0)$, with the terms of highest transcendentality 
of the HEL of the projection of the two-loop amplitude on the tree 
amplitude~\cite{DelDuca:2001gu}.

Equating the coefficients of the single logarithm of 
\eqns{exp2loopu}{eq:a42ubds}, 
we obtain the MSYM two-loop trajectory~\cite{Kotikov:2002ab} in the DRED
scheme,
\beq
\alpha^{(2)} = - {\pi^2\over 3\eps} - 2\zeta_3 - {4\pi^4\over 45}\eps
+ (6\pi^2\zeta_3 + 82\zeta_5)\eps^2 + \ord(\eps^3)\, ,\label{eq:tworegge}
\eeq
which is in agreement, to $\ord(\eps^0)$, with the terms of highest 
transcendentality of the QCD two-loop trajectory~\cite{DelDuca:2001gu}.

Equating the coefficients of the constant term of \eqns{exp2loopu}{eq:a42ubds},
we obtain the MSYM two-loop coefficient function,
\bea
C^{(2)}_{MSYM} &=& 
\frac{2}{\eps^4} - {5\pi^2\over 6}\frac{1}{\eps^2} 
- \frac{\zeta_3}{\eps} - {11\over 72}\pi^4 \nn\\ &+& 
\left( {\pi^2\over 6}\zeta_3 - 41\zeta_5 \right) \eps 
- \left( {95\over 2}\zeta_3^2 + {113\pi^6\over 504} \right)
\eps^2 + \ord(\eps^3)\, .
\label{eq:2loopif}
\eea
The evaluation of \eqn{eq:2loopif} has been performed also in the $s$ channel
through \eqns{exp2loops}{eq:a42sbds}, thus checking the consistency of 
\eqns{elasschan}{elasuchan} at two loops.

At three loops, we take the direct calculation of the exact MSYM amplitude,
provided in Ref.~\cite{Bern:2005iz} to $\ord(\eps^0)$,
and evaluate the HEL in the $u$ and $s$ physical channels. We find
\bea
m_{4_{MSYM}}^{u(3)} &=& 
- \frac{32}{3\eps^6} + \frac{16 L}{\eps^5} -
\left( 8L^2 - {20\pi^2\over 3}\right) \frac{1}{\eps^4} +
\left[ {4\over 3} L^3 - 6\pi^2 L + 8\zeta_3\right] \frac{1}{\eps^3}
\nn\\ &+&
\left[ {4 \pi^2\over 3} L^2 - 4\zeta_3 L + {181\over 405}\pi^4 \right] 
\frac{1}{\eps^2} -
\left( {26\over 135}\pi^4 L + {112\over 27}\zeta_3\pi^2 - {952\over 3}\zeta_5
\right) \frac{1}{\eps}
\nn\\ &-&
{2\pi^4\over 45} L^2 
- \left( 484\zeta_5 + {196\pi^2\over 9}\zeta_3 \right) L
+ {3284\over 9}\zeta_3^2 + {88747\pi^6\over 51030} + \ord(\eps)\, ,
\label{eq:a43ubds}\\
m_{4_{MSYM}}^{s(3)} &=& m_{4_{MSYM}}^{u(3)}
- \frac{16i\pi}{\eps^5} + \frac{8\pi^2 + 16 L i\pi}{\eps^4}
- \left[4\pi^2 L + \left( 4 L^2 - {22\pi^2 \over 3} \right) i\pi\right]
\frac{1}{\eps^3} \nn\\ && \qquad\qquad -
\left[ {4\pi^4\over 3} + \left( {8 \pi^2\over 3} L - 4\zeta_3 \right) i\pi
\right] \frac{1}{\eps^2}
+ {26\over 135}\frac{i\pi^5}{\eps} \nn\\
&& \qquad\qquad + {2\pi^6\over 45} 
+ \left( {4\pi^4\over 45} L + {196\over 9}\zeta_3\pi^2
+ 484 \zeta_5 \right) \pi i + \ord(\eps)\, .
\label{eq:a43sbds}
\eea
The coefficients of the double logarithm of \eqns{exp3loopu}{eq:a43ubds}
provide a cross check of the two-loop trajectory $\alpha^{(2)}$, while
comparing the coefficients of the single logarithm we obtain
the MSYM three-loop trajectory,
\beq
\alpha^{(3)} = {22\over 135}{\pi^4\over\eps} + {20\pi^2\over 9}\zeta_3 +
16\zeta_5 + \ord(\eps) \, ,\label{eq:treregge}
\eeq
in agreement with Ref.~\cite{Bartels:2008ce,Drummond:2007aua}.
The trajectory was also implicitly evaluated in Ref.~\cite{Naculich:2007ub}.
Equating the coefficients of the constant term of \eqns{exp3loopu}{eq:a43ubds}
we find the MSYM three-loop coefficient function,
\bea
C^{(3)}_{MSYM} &=& -\frac{4}{3\eps^6} + \frac{2\pi^2}{3}{1\over\eps^4} 
+ {217\pi^4\over 810}\frac{1}{\eps^2} +
\left( - {11\pi^2\over 27}\zeta_3 + {224\over 3}\zeta_5 \right) \frac{1}{\eps} 
\nn\\ &+& \left( {796\over 9}\zeta_3^2 + {211861\pi^6\over 408240} \right)
+ \ord(\eps)\, .
\label{eq:3loopif}
\eea
The evaluation of \eqn{eq:3loopif} has been performed also in the $s$ channel
through \eqns{exp3loops}{eq:a43sbds}, thus checking the consistency of 
\eqns{elasschan}{elasuchan} at three loops.

\section{The Bern-Dixon-Smirnov ansatz}
\label{sec:bds}

The BDS ansatz prescribes that the $n$-point MHV amplitude
be written as,
\bea
m_n &=& m_n^{(0)} \left[ 1 + \sum_{L=1}^\infty a^L M_n^{(L)}(\eps) \right] 
\nn\\ &=& m_n^{(0)} 
\exp\left[ \sum_{l=1}^\infty a^l \left( f^{(l)}(\eps) 
M_n^{(1)}(l\eps) + Const^{(l)} + E_n^{(l)}(\eps)\right)\right]\, 
,\label{eq:bds1}
\eea
where
\beq
a = {2\gs^2 N\over (4\pi)^{2-\eps}} e^{-\gamma\eps}
\eeq
is the 't-Hooft gauge coupling, and with
%
%
\beq
f^{(l)}(\eps) = f^{(l)}_0 + \eps f^{(l)}_1 + \eps^2 f^{(l)}_2\, 
,\label{eq:flfunct}
\eeq
where $f^{(1)}(\eps)=1$, and
$f^{(l)}_0$ is proportional to the $l$-loop cusp anomalous 
dimension, $\hat{\gamma}_K^{(l)} = 4f^{(l)}_0$ and $f^{(l)}_1$
is related to another quantity, ${\cal G}_0^{(l)} = 2f^{(l)}_1/l$,
which enters the IR Sudakov form factor and accounts for virtual divergences
which are not simultaneously soft and 
collinear~\cite{Magnea:1990zb,Sterman:2002qn}. In \eqn{eq:bds1},
$Const^{(l)}$ are constants, and $E_n^{(l)}(\eps)$ are $\ord(\eps)$ 
contributions, with $Const^{(1)}=0$ and $E_n^{(1)}(\eps)=0$,
and $M_n^{(L)}(\eps)$ is the $L$-loop colour-stripped
amplitude rescaled by the tree amplitude. In the convention and notation 
of \eqn{elasexpand}, the four-point amplitude is given by,
\beq
a^L M_4^{(L)}(\eps) = \left( \frac{a}{2G(\eps)}\right)^L
\left( {\mu^2\over -t}\right)^{L\eps} m_4^{(L)}(\eps)\, ,\label{eq:ourm}
\eeq
with
\beq
G(\eps) = {e^{-\gamma\eps}\ \Gamma(1-2\epsilon)\over
\Gamma(1+\epsilon)\, \Gamma^2(1-\epsilon)} = 1 + \ord(\eps^2)\, ,
\eeq
Thus, using the rescaled coupling (\ref{eq:nrescale}), the BDS ansatz
(\ref{eq:bds1}) for the four-point amplitude becomes
\bea
m_4 &=& m_4^{(0)} \left[ 1 + \sum_{L=1}^\infty {\bar\gs}^{2L}(t) 
m_4^{(L)}(\eps) \right] \nn\\ &=& m_4^{(0)} 
\exp\left[ \sum_{l=1}^\infty {\bar\gs}^{2l}(t) \left( 2G(\eps)\right)^l 
\left( f^{(l)}(\eps) {m_4^{(1)}(l\eps)\over 2G(l\eps)}
+ Const^{(l)} + \ord(\eps)\right)\right]\, .\label{eq:bdsddg}
\eea
\eqn{eq:bdsddg} applies equally well
in either the $s$ or the $u$ channels.
%
%
Substituting the HEL one-loop amplitude (\ref{exp1loop}) in
\eqn{eq:bdsddg} and comparing
with the expansion (\ref{elasexpand}) of the high-energy 
factorization (\ref{elasschan}) or (\ref{elasuchan}), 
we see that the coefficient of the
single logarithm allows us to read off the value of the gluon trajectory,
\bea
\alpha^{(1)} &=& {2\over\eps}\ G(\eps) \nn\\
\alpha^{(2)} &=& {2\over\eps}\ G^2(\eps) f^{(2)}(\eps) \label{eq:alphabds}\\
\alpha^{(3)} &=& {8\over 3\eps}\ G^3(\eps) f^{(3)}(\eps)\, ,\nn
\eea
and in general
\beq
\alpha^{(l)} = {2^l\over l\eps}\ G^l(\eps) f^{(l)}(\eps)\, .\label{eq:alphagen}
\eeq
From \eqn{eq:alphagen}, we see that
only the first two terms of the $f^{(l)}(\eps)$ function (\ref{eq:flfunct})
enter the evaluation of the Regge trajectory.
Dropping $G(\eps)$, which does not contribute in \eqn{eq:alphabds}
to $\ord(\eps^0)$, and using the $f^{(2)}$ and $f^{(3)}$
functions~\cite{Bern:2005iz},
\bea
f^{(2)}(\eps) &=& - \zeta_2 - \zeta_3\eps - \zeta_4\eps^2\, ,\nn\\
f^{(3)}(\eps) &=& {11\over 2} \zeta_4 + (6\zeta_5 + 5\zeta_2\zeta_3)\eps
+ (c_1\zeta_6 + c_2\zeta_3^2)\eps^2\, ,\label{eq:ffunct} 
\eea
we see that \eqn{eq:alphabds} agrees with \eqns{eq:tworegge}{eq:treregge}
to $\ord(\eps^0)$. The coefficients $c_1, c_2$ are unknown, but they
do not enter the evaluation of the Regge trajectory.

Through the iterative structure of the MSYM amplitudes, it is possible
to express the coefficient functions at a given loop in terms of
the coefficient functions at a lower number of loops. Using \eqn{eq:bdsddg},
the iterative structure of the two-loop four-point MSYM 
amplitude~\cite{Anastasiou:2003kj,Bern:2004cz} is given by
\beq
m_4^{(2)}(\eps) = {1\over 2} \left[m_4^{(1)}(\eps)\right]^2
+ {2\,G^2(\eps)\over G(2\eps)} f^{(2)}(\eps)\, m_4^{(1)}(2\eps) 
+ 4\, Const^{(2)} + \ord(\eps)\, ,\label{eq:ite2}
\eeq
with $Const^{(2)}= -\zeta_2^2/2$, and where the one-loop amplitude must be
known to $\ord(\eps^2)$. Using the two-loop factorization,
in either the $u$ (\ref{exp2loopu}) or $s$ channels (\ref{exp2loops}),
we find
\beq
C^{(2)}_{MSYM}(\eps) = {1\over 2} \left[ C^{(1)}_{MSYM}(\eps)\right]^2
+ {2\,G^2(\eps)\over G(2\eps)} f^{(2)}(\eps)\, C^{(1)}_{MSYM}(2\eps) 
+ 2\, Const^{(2)}
+ \ord(\eps)\, ,\label{eq:ifite2}
\eeq
where the one-loop coefficient function is needed to $\ord(\eps^2)$.
\eqn{eq:ifite2} agrees with \eqn{eq:2loopif} to $\ord(\eps^0)$.

Similarly, the iterative structure of the three-loop four-point MSYM amplitude 
is~\cite{Bern:2005iz},
\beq
m_4^{(3)}(\eps) = - {1\over 3} \left[m_4^{(1)}(\eps)\right]^3
+ m_4^{(2)}(\eps)\, m_4^{(1)}(\eps)
+ {4\,G^3(\eps)\over G(3\eps)} f^{(3)}(\eps)\, m_4^{(1)}(3\eps) 
+ 8\, Const^{(3)} + \ord(\eps)\, ,\label{eq:ite3}
\eeq
where $m_4^{(1)}(\eps)$ and $m_4^{(2)}(\eps)$ must be known to
$\ord(\eps^4)$ and $\ord(\eps^2)$, respectively, and with
\beq
Const^{(3)} = \left( {341\over 216} + {2\over 9} c_1\right) \zeta_6
+ \left( -{17\over 9} + {2\over 9} c_2\right) \zeta_3^2\, .\label{eq:cost3}
\eeq
Using the three-loop factorization, in either the $u$ (\ref{exp3loopu}) 
or $s$ channels (\ref{exp3loops}), we obtain the three-loop coefficient function
\bea
C^{(3)}_{MSYM}(\eps) &=& - {1\over 3} \left[C^{(1)}_{MSYM}(\eps)\right]^3
+ C^{(1)}_{MSYM}(\eps)\, C^{(2)}_{MSYM}(\eps) \nn\\
&+& {4\,G^3(\eps)\over G(3\eps)} f^{(3)}(\eps)\, C^{(1)}_{MSYM}(3\eps)
+ 4\, Const^{(3)} + \ord(\eps)\, .\label{eq:ifite3}
\eea
The coefficients $c_1, c_2$ cancel when \eqns{eq:ffunct}{eq:cost3} are used
in \eqns{eq:ite3}{eq:ifite3}.
Using the two-loop coefficient function to $\ord(\eps^2)$ (\ref{eq:2loopif}),
and the one-loop coefficient function to $\ord(\eps^4)$ (\ref{eq:ifonel}),
we see that \eqn{eq:ifite3} is in agreement with \eqn{eq:3loopif} to 
$\ord(\eps^0)$.

\section{Conclusions}
\label{sec:concl}

The iterative structure and the exponentiated form of the MSYM amplitudes
in the MHV configuration offer a useful computational lab
to test high-energy factorization in the MSYM, and allow the
derivation of some of the relevant quantities in the high energy limit.
Using the MSYM two- and three-loop four-point amplitudes, 
we have tested the high-energy factorization of the colour-stripped amplitude. 
In particular, we have shown that it is valid beyond NLL accuracy, and we have 
verified that the factorization formulae (\ref{elasschan}) and 
(\ref{elasuchan}) hold at three-loop accuracy. Accordingly, we have derived the
three-loop Regge trajectory (\ref{eq:treregge}), as well as the 
two-loop (\ref{eq:2loopif}) and three-loop coefficient 
functions (\ref{eq:3loopif}). 

The three-loop Regge 
trajectory is one of the building blocks for deriving a BFKL evolution equation
at next-to-next-to-leading logarithmic (NNLL) accuracy. The others are
the tree vertex for the emission of 
three gluons along the ladder~\cite{Del Duca:1999ha,Antonov:2004hh},
the two-loop vertex for the emission of a gluon along the ladder, and
the one-loop vertex for the emission of two gluons along the ladder.
In addition, if one wants to compute jet cross sections at a matching accuracy,
the NNLO impact factors must be determined. The ingredients for that
are the two-loop coefficient function evaluated here, as well as
the tree coefficient function for the emission of three 
gluons~\cite{Del Duca:1999ha,Antonov:2004hh}, 
and the one-loop coefficient function for the emission of two gluons.

Furthermore the two-loop and three-loop Regge trajectories and coefficient functions,
together with the vertices for the emission of one or more gluons 
along the ladder which were not analysed in this work, can be used to build MSYM 
two-loop and three-loop amplitudes with a larger number of legs in the high-energy limit.
This may serve as a powerful check on the structure of high multiplicity MSYM amplitudes.

\section*{Acknowledgements}

We thank Lance Dixon for useful suggestions, and Alejandro Daleo
for help with the code of Ref.~\cite{Maitre:2005uu}. 
This work was partly supported by MIUR 
under contract 2006020509$_0$04 and by the EC Marie-Curie Research 
Training Network ``Tools and Precision
Calculations for Physics Discoveries at Colliders''
under contract MRTN-CT-2006-035505 .

\end{document}